\documentclass[a4paper]{jpconf}
\usepackage{graphicx}
\usepackage{amsmath}
\usepackage{cite}
\begin{document}
\title{Mid-rapidity D-meson production in pp, Pb--Pb and p--Pb collisions at the LHC}

\author{Jeremy Wilkinson for the ALICE Collaboration}

\address{Physikalisches Institut, Ruprecht-Karls-Universit\"{a}t Heidelberg, Heidelberg, Germany}

\ead{jwilkinson@physi.uni-heidelberg.de}

\begin{abstract}
We present the recent results for D-meson production measured by the ALICE collaboration in pp collisions at $\sqrt{s}=7$~TeV, Pb--Pb collisions at $\sqrt{s_\mathrm{NN}}=2.76$~TeV and p--Pb collisions at $\sqrt{s_\mathrm{NN}}=5.02$~TeV. 

\end{abstract}
\section{Introduction} \label{sec:intro}	
Measurements of heavy-flavour production at a variety of energies and in various colliding systems offer a unique probe into the properties of the quark-gluon plasma (QGP) formed in heavy-ion collisions. Due to their large mass, charm and beauty quarks are produced in the early stages of the collision through hard partonic interactions. This means that they maintain their identity throughout the full evolution of the collision, thereby serving as a tagged probe.

The ALICE experiment studies heavy-flavour production in part through the reconstruction of D mesons produced in the collisions. This is performed in three collision systems, each giving different insights into the overall mechanisms of charm production. Proton-proton collisions serve as a fundamental test of perturbative quantum chromodynamic (pQCD) calculations, as well as providing an essential reference spectrum with which heavy-ion results can be compared; Pb--Pb collisions give the possibility of studying the properties of strongly-interacting matter at high temperatures and densities; and p--Pb collisions allow us to disentangle cold nuclear effects (such as $k_\mathrm{T}$ broadening, the nuclear modification of parton distribution functions, and parton saturation at low $x$) from hot nuclear matter effects arising in the medium.

The medium's properties can be studied using two of the experimental observables extracted from heavy-ion results, namely the  nuclear modification factor $R_\mathrm{AA}$ and the elliptic flow parameter $v_2$. The $R_\mathrm{AA}$ measures modifications of the transverse momentum distributions of D mesons, and is sensitive to the energy loss of charm quarks due to e.g. gluon bremsstrahlung and elastic collisions with other constituents of the medium. It is defined as
\begin{equation}
R_\mathrm{AA}(p_\mathrm{T}) = \frac{1}{\left<N_\mathrm{coll}\right>}\frac{\mathrm{d}N_\mathrm{AA}/\mathrm{d}p_\mathrm{T}}{\mathrm{d}N_\mathrm{pp}/\mathrm{d}p_\mathrm{T}}, \label{RAA}
\end{equation}
where $\left<N_\mathrm{coll}\right>$ is the number of binary collisions, $N_\mathrm{AA}$ is the measured yield in Pb--Pb collisions and $N_\mathrm{pp}$ is the yield in pp collisions at the same collision energy. It is expected that $R_\mathrm{AA}$ should be equal to unity if there are no nuclear effects present. In addition, the energy loss depends on colour charge and the quark-mass-dependent dead cone effect~\cite{deadcone}. Due to this, there is an expected hierarchy of suppression between different parton species. This cannot be measured directly, but will also alter the production of their decay products. It is expected, therefore, that $R_\mathrm{AA}(\pi) < R_\mathrm{AA}(\mathrm{D}) < R_\mathrm{AA}(\mathrm{B})$.

The elliptic flow $v_2$ is a measure of the momentum anisotropy of the particles produced in an interaction. Such anisotropy occurs especially in semi-peripheral heavy-ion collisions due to the overlap region of the nuclei being spatially anisotropic. This can be expressed with a Fourier expansion of the distribution of produced particles as a function of the azimuthal angle $\varphi$ relative to the reaction plane $\Psi_\mathrm{RP}$:
\begin{equation}
\frac{\mathrm{d}N}{\mathrm{d}(\varphi-\Psi_\mathrm{RP})} = \frac{N_0}{2\pi}(1+2v_1\cos(\varphi-\Psi_\mathrm{RP})+2v_2\cos(2(\varphi-\Psi_\mathrm{RP}))+...),
\label{v2}
\end{equation}
with $v_n$ representing the different harmonics of the expansion. The reaction plane is defined so as to contain both the impact parameter and the beam axis. At low and intermediate $p_\mathrm{T}$, the value of $v_2$ offers an insight into the level of thermalisation of charm quarks in the medium, while at higher $p_\mathrm{T}$ it allows us to study the path-length dependence of heavy-quark energy loss.

In these proceedings, we will discuss the ALICE measurements of D-meson production in pp collisions at $\sqrt{s}=7$~TeV, Pb--Pb collisions at $\sqrt{s_\mathrm{NN}}=2.76$~TeV, and p--Pb collisions at $\sqrt{s_\mathrm{NN}}=5.02$~TeV. 

\section{Analysis strategy}
The analysis of D mesons reconstructed inside the ALICE central barrel makes use of the detector's excellent tracking, vertexing and particle identification (PID) capabilities.

For this analysis, we fully reconstruct D mesons (and their respective charge conjugates) in their hadronic decay channels, with branching ratios $\cal B$ and mean proper decay lengths $c\tau$ as follows~\cite{PDGreview}: \begin{align*}
&\mathrm{D}^0\rightarrow{}\mathrm{K}^{-}\pi^+ ({\cal B} = 3.88 \pm 0.05\%, c\tau = 123\mu\mathrm{m});\\[-0.3pc]
&\mathrm{D}^+\rightarrow{}\mathrm{K}^{-}\pi^+\pi^+ ({\cal B} = 9.13 \pm 0.19\%, c\tau \approx 312\mu\mathrm{m});\\[-0.3pc]
&\mathrm{D}^{*+}\rightarrow{}\mathrm{D}^0\pi^+ ({\cal B} = 67.7 \pm 0.5\%) \rightarrow \mathrm{K}^-\pi^+\pi^+;\\[-0.3pc]
&\mathrm{D}_{\mathrm{s}}^+\rightarrow\varphi\pi^+\rightarrow\mathrm{K}^+\mathrm{K}^-\pi^+ ({\cal B} = 2.28 \pm 0.12\%, c\tau \approx 150\mu\mathrm{m}).\end{align*}

The analysis of D-meson production mainly makes use of four detector systems: the Inner Tracking System (ITS), the VZERO detector, the Time Projection Chamber (TPC) and the Time-of-Flight detector (TOF). The ITS, formed of a series of silicon pixel, strip and drift detectors, offers high-resolution tracking and vertexing at mid-rapidity ($|\eta| <$ 0.9). The VZERO detector is a scintillator detector covering the pseudorapidity ranges $2.8 < \eta < 5.1$ (VZEROA) and $-3.7 < \eta < -1.7$ (VZEROC), and is used for triggering and centrality determination. The TPC is a gas detector, which surrounds the ITS in the region $|\eta| <$ 0.9 and offers further tracking as well as PID via measurements of specific energy loss. Finally, the TOF detector offers PID at mid-rapidity via measurements of particles' times of flight.

As D mesons have a non-zero lifetime, they typically travel some distance from the interaction point before decaying.
This allows us to make topological selections of possible decay daughters based on e.g. the impact parameter (the distance of closest approach between the daughter track and the primary vertex) and the pointing angle (the angle between the meson flight line and its reconstructed momentum).
In the case of the D$^{*+}$ meson, the lifetime is eight orders of magnitude shorter than the other species, making a direct topological selection impossible. Instead we pair a D$^0$ candidate with a $\pi^\pm$ at the primary vertex to reconstruct these. 

In tandem with the topological selections, we also use the PID capabilities of the TPC and TOF to identify the daughter tracks based on the measured responses in these detectors. Accepted candidates are then used to fill $p_\mathrm{T}$-binned invariant mass spectra, from which the raw yields are extracted using a fit function composed of a Gaussian function for the signal and an exponential function for the background. The mean of the signal peak is typically found to be compatible with to the PDG mass value~\cite{PDGreview}. The width of the peak, which for D$^0$ and D$^+$ varies between {\raise.17ex\hbox{$\scriptstyle\mathtt{\sim}$}}6 and {\raise.17ex\hbox{$\scriptstyle\mathtt{\sim}$}}20~MeV/$c^2$, is well reproduced by Monte Carlo (MC) simulations. These yields are then corrected for their detection efficiencies, which are determined using MC simulations of the specific collision system.

Finally, it is necessary to subtract feed-down due to the decays of B mesons produced in the collisions. This subtraction is performed using FONLL~\cite{FONLL} calculations, the reconstruction and selection efficiency of D mesons from B feed-down, and a hypothesis of the $R_\mathrm{AA}$ ($R_\mathrm{pPb}$) of D mesons arising from B-meson decays in Pb--Pb (p--Pb) collisions.

\section{Results in pp collisions} \label{sec:ppresults}

\begin{figure}[h!t]
\centering
\begin{minipage}[t]{13.9pc}\vspace{0px}
\includegraphics[width=13.9pc]{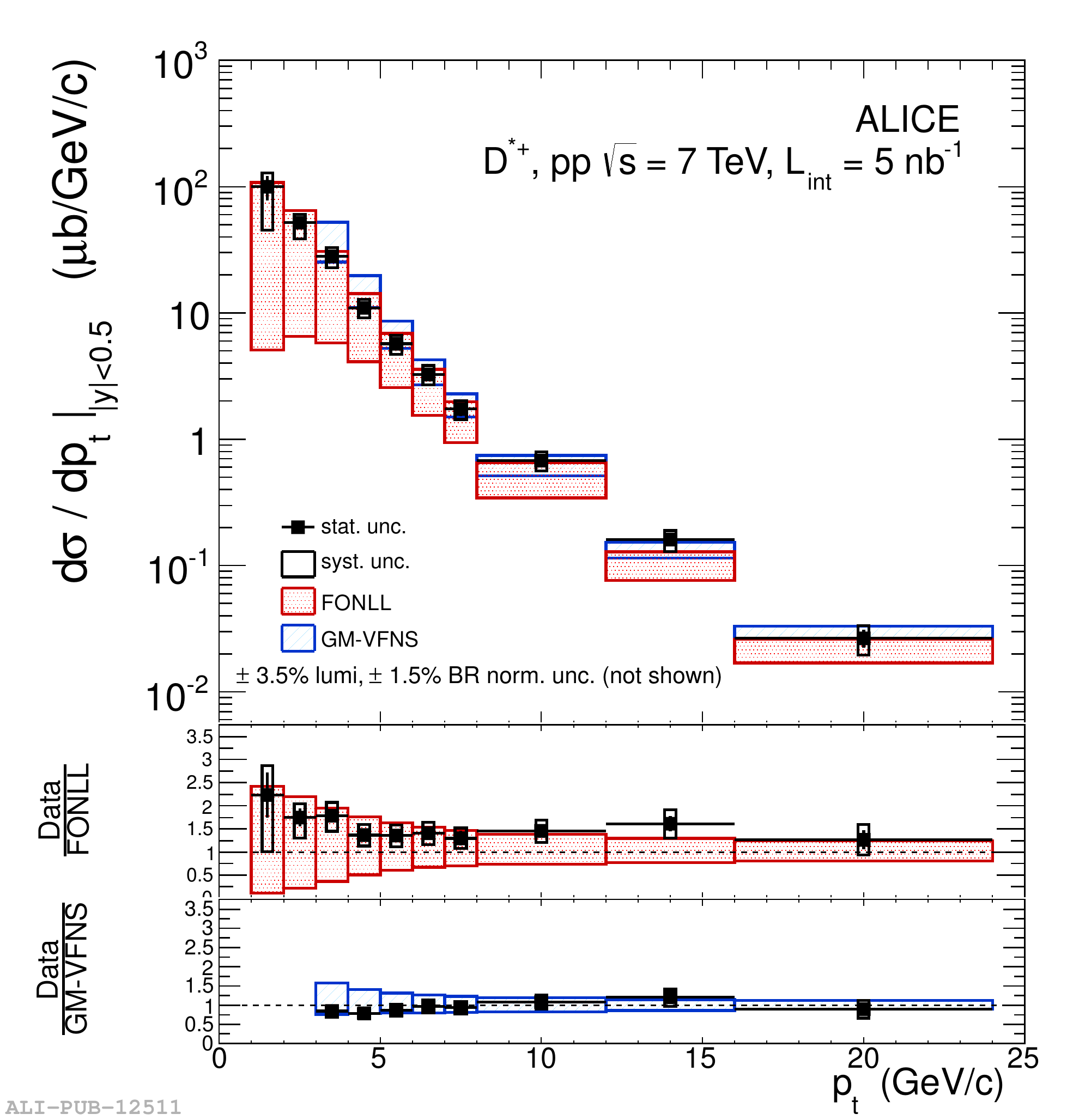}
\caption{\label{fig:diffcrosssectionspp7TeV}$p_\mathrm{T}$-differential cross section of D$^{*+}$ production in pp collisions at $\sqrt{s}=7$~TeV, compared with results from FONLL~\cite{FONLL} and GM-VFNS~\cite{GM-VFNS} pQCD calculations~\cite{ALICEpp7TeVDmesons}.}
\end{minipage}\hspace{2pc}%
\begin{minipage}[t]{18.3pc}\vspace{-1.8px}
\includegraphics[width=18.3pc]{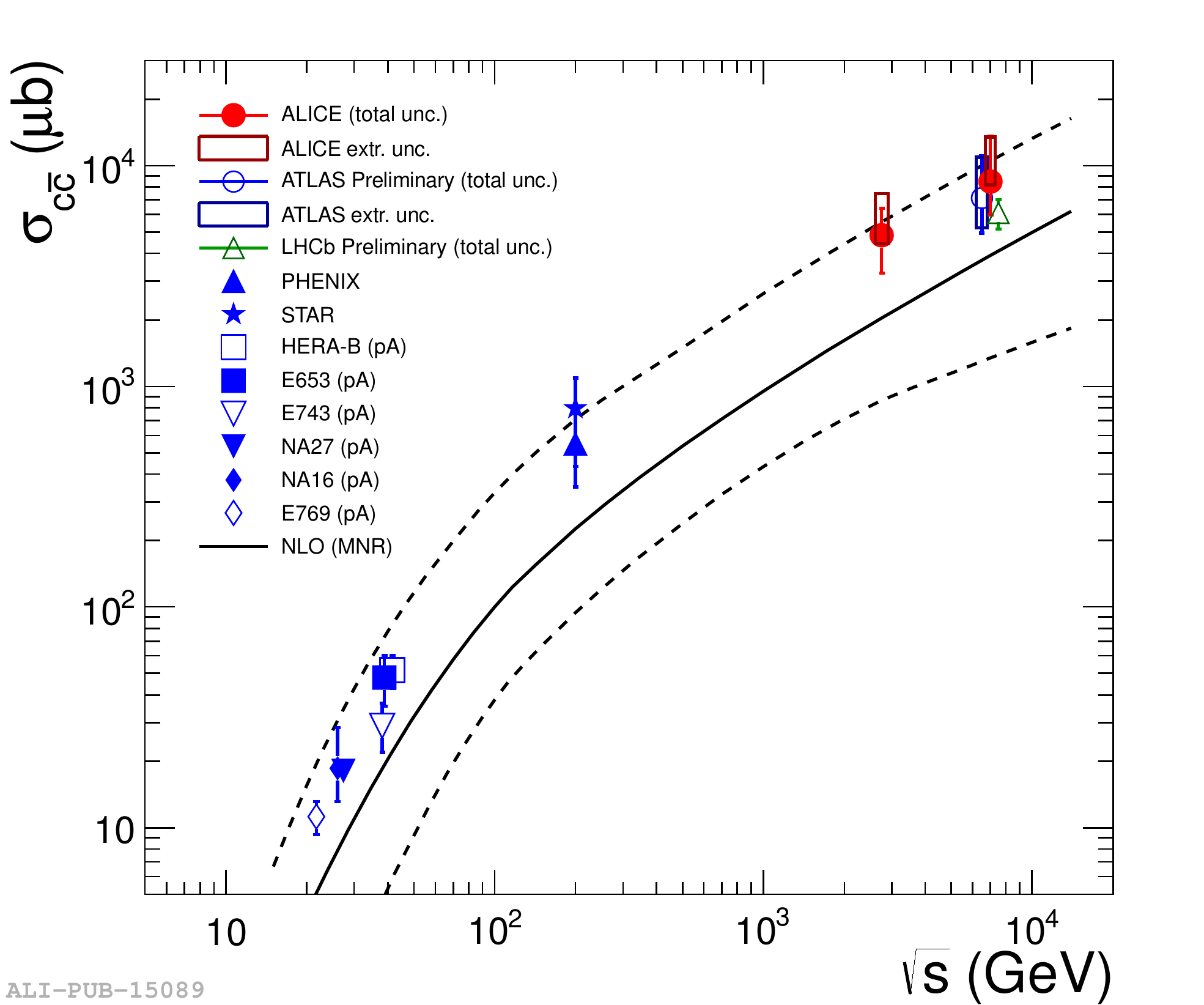}\vspace{-7px}
\caption{\label{fig:totalcharmcrosssectionpp}Total charm cross section measured in pp collisions with ALICE as a function of $\sqrt{s}$, in comparison with other experiments and MNR~\cite{MNR} calculations~\cite{ALICEpp276TeVDmesons}.}
\end{minipage} 
\end{figure}

The analysis in pp collisions at $\sqrt{s}=7$~TeV was performed using a data sample taken in 2010 with a minimum-bias trigger. The analysed data corresponded to an integrated luminosity of 5~nb$^{-1}$. This measurement was made at mid-rapidity, in the region $1 < p_\mathrm{T} < 24$~GeV/$c$. The differential cross sections were reported in~\cite{ALICEpp7TeVDmesons}, and the result for D$^{*+}$ production is shown in figure \ref{fig:diffcrosssectionspp7TeV}. 
The measured cross sections are described within uncertainties by FONLL~\cite{FONLL} and GM-VFNS~\cite{GM-VFNS} pQCD calculations, with the experimental results matching the upper end of the uncertainty bands in the FONLL calculations.

Further measurements were also taken for pp collisions at $\sqrt{s}=2.76$~TeV, which was the centre-of-mass energy used for Pb--Pb collisions. These results were reported in~\cite{ALICEpp276TeVDmesons}. Since the statistics collected at this energy were fewer than those recorded at $\sqrt{s}=7$~TeV, these were not used  as the pp reference for the Pb--Pb results, instead being used mainly to verify the validity of an energy scaling of the 7~TeV results to $\sqrt{s}=2.76$~TeV using FONLL~\cite{energyscaling}.

In addition, the results for 7~TeV and 2.76~TeV pp collisions were extrapolated to full phase space using FONLL calculations, in order to extract the total c$\bar{\mathrm{c}}$ cross section in pp at these energies. The results of these extrapolations are shown in figure \ref{fig:totalcharmcrosssectionpp}, alongside results from other experiments and next-to-leading-order pQCD calculations using the MNR framework~\cite{MNR}. The result at 7~TeV is in agreement with those from the ATLAS and LHCb collaborations, and both of the ALICE points follow the trends exhibited by both the NLO predictions and other experimental results spanning three orders of magnitude in energy. One thing to note here is that all of the results lie within the upper part of the uncertainty band of the predictions; this is possibly due to the central calculation overestimating the bare charm quark mass ($m_\mathrm{c} = 1.5$~GeV/$c^2$).

\section{Results in Pb--Pb collisions} \label{sec:pbpbresults}
The Pb--Pb data at $\sqrt{s_\mathrm{NN}}=2.76$~TeV were taken in 2010 and 2011, using both minimum-bias and centrality-triggered data. In 2010, a data sample of 2~$\mu\mathrm{b}^{-1}$ was taken with a minimum-bias trigger, while in 2011 an online selection based on VZERO was used to enhance the sample of central ($28~\mu\mathrm{b}^{-1}$) and mid-central ($6~\mu\mathrm{b}^{-1}$) collisions. In order to provide the reference spectrum at $\sqrt{s}=2.76$~TeV, the high-statistics 7~TeV data were scaled to 2.76~TeV using FONLL, as discussed in section~\ref{sec:ppresults}. The nuclear modification factors of D$^0$, D$^+$ and D$^{*+}$ mesons using the 2010 data sample were published in~\cite{alicepbpbraa}, and the $v_2$ results for these from the 2011 data sample were published in~\cite{alicepbpbv2}.

\begin{figure}[h!t]
\centering
\begin{minipage}[t]{13.8pc}\vspace{0px}
\includegraphics[width=13.8pc]{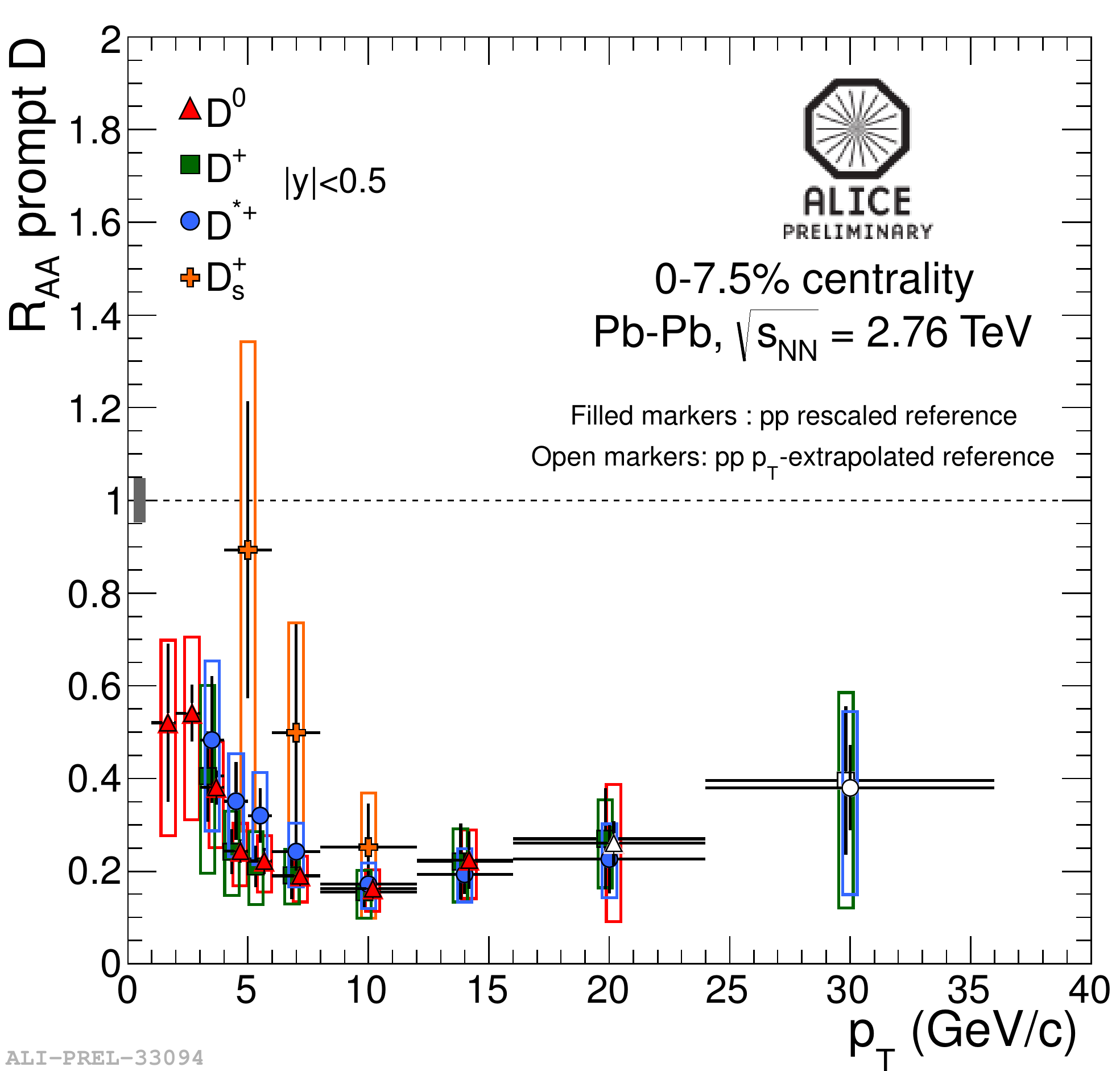}
\caption{\label{fig:centralraa}  $R_\mathrm{AA}$ of D$^0$, D$^+$, D$^{*+}$, and D$_\mathrm{s}^+$ mesons in central Pb--Pb collisions at $\sqrt{s_\mathrm{NN}}=2.76$~TeV.}
\end{minipage}\hspace{2pc}%
\begin{minipage}[t]{17.8pc}\vspace{-4.7px}
\includegraphics[width=17.8pc]{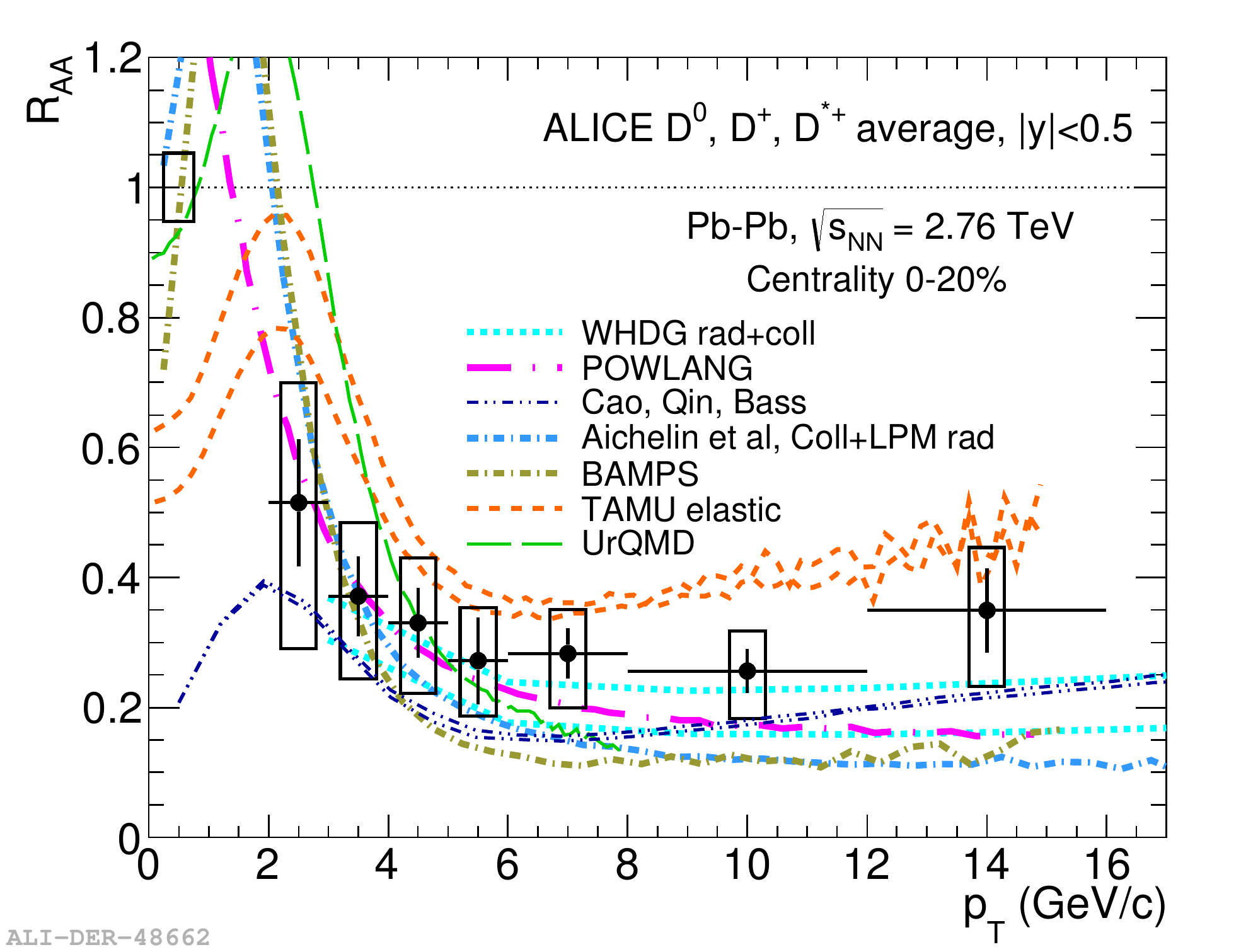}\vspace{-0.9px}
\caption{\label{fig:centralraavsmodels} Average $R_\mathrm{AA}$ for D$^0$, D$^+$, and D$^{*+}$ mesons, compared with model predictions~\cite{WHDG,POWLANG,caoqinbass,aichelin1,aichelin2,BAMPS,TAMU,UrQMD1,UrQMD2}.}
\end{minipage} 
\end{figure}

The $R_\mathrm{AA}$ for the 7.5\% most-central Pb--Pb collisions was measured for all four D-meson species in different $p_\mathrm{T}$ regions spanning $1 < p_\mathrm{T} < 36$~GeV/$c$. The measurements for each species are shown in figure~\ref{fig:centralraa}. In figure~\ref{fig:centralraavsmodels}, the average $R_\mathrm{AA}$ of D$^0$, D$^+$ and D$^{*+}$ mesons in the centrality range 0-20\%, taken from~\cite{alicepbpbraa}, is compared with different model calculations implementing in-medium parton energy loss.

The results for D$^0$, D$^+$ and D$^{*+}$ agree with one another within uncertainties, and display a particularly high level of suppression at $p_\mathrm{T} > 6$~GeV/$c$. The $R_\mathrm{AA}$ of D$_\mathrm{s}^+$, measured in the range $4 < p_\mathrm{T} < 12$~GeV/$c$, is also displayed. At present, the uncertainties are too large to conclude whether or not there is a lower level of suppression for D$_\mathrm{s}^+$ at low to intermediate $p_\mathrm{T}$ in comparison with the other D mesons. Such lifting of the strangeness suppression in Pb--Pb compared to pp is theorised to possibly happen due to charm recombination in the medium~\cite{charmrecombin1,charmrecombin2}.

\begin{figure}[h!t]
\centering
\begin{minipage}{14pc}
\includegraphics[width=14pc]{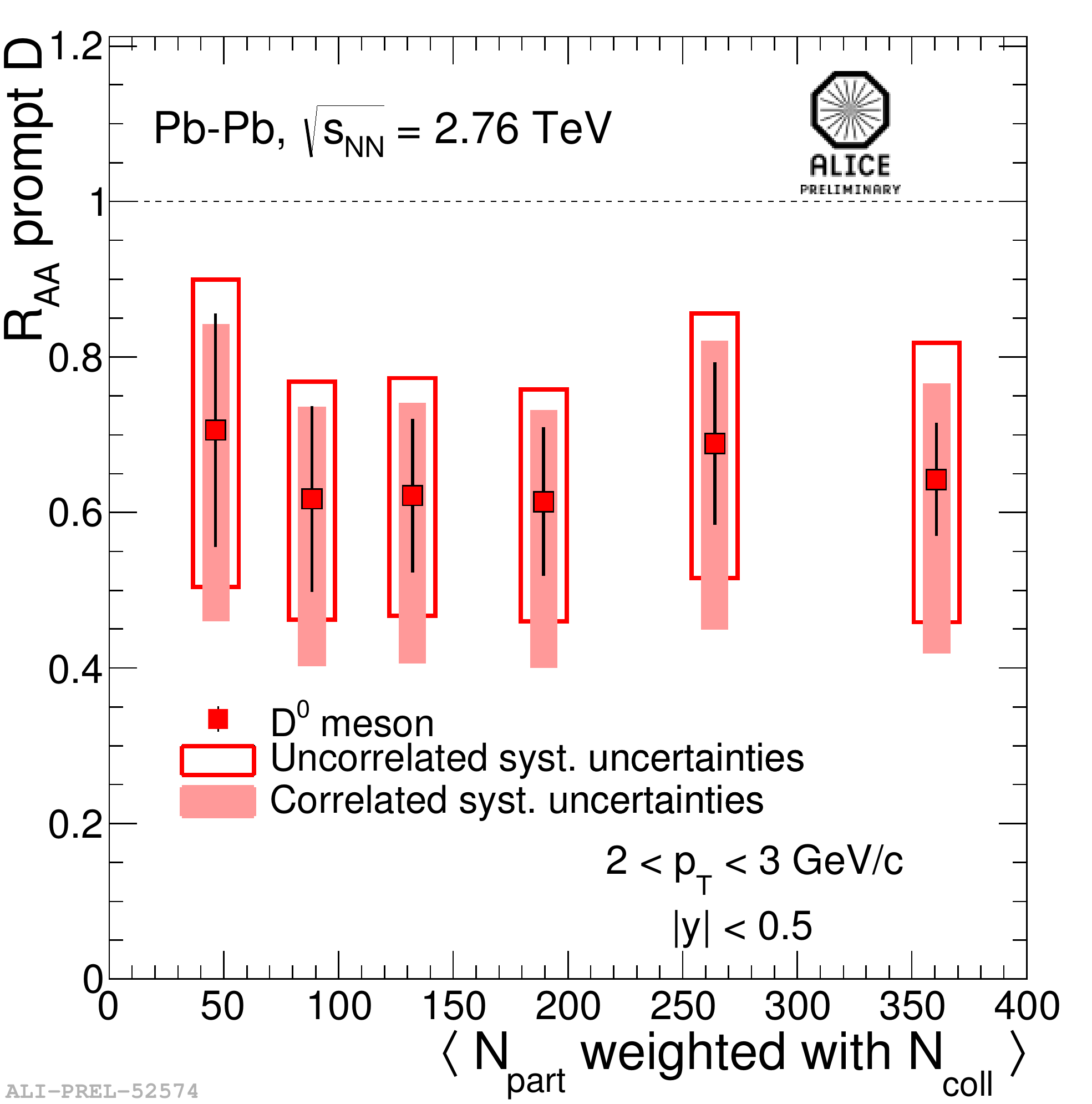}
\end{minipage}\hspace{2pc}%
\begin{minipage}{14pc}
\includegraphics[width=14pc]{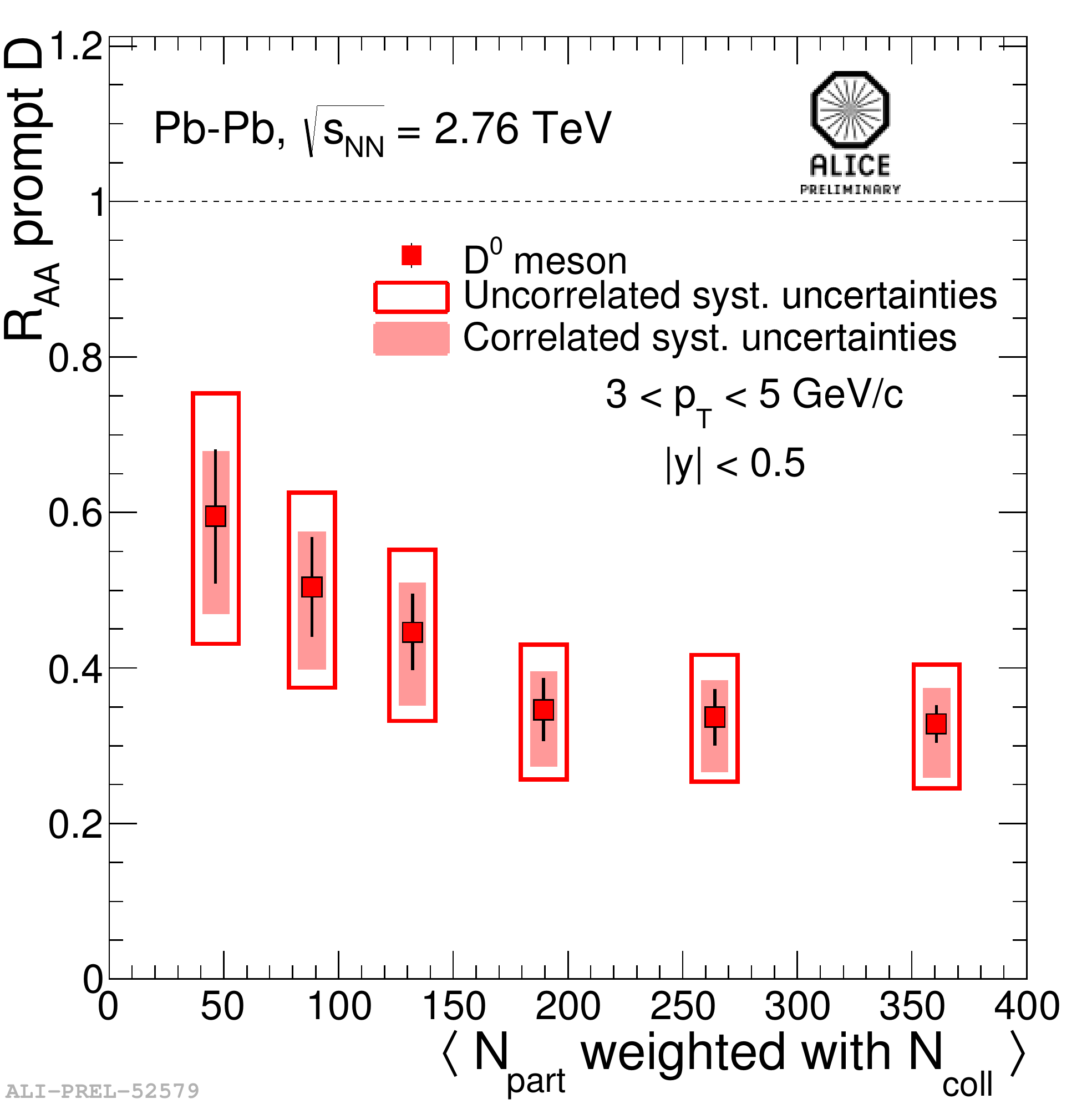}
\end{minipage} 
\caption{\label{fig:D0raavscent1} \label{fig:D0raavscent2}  $R_\mathrm{AA}$ as a function of centrality for D$^0$ mesons in Pb--Pb collisions at $\sqrt{s_\mathrm{NN}}=2.76$~TeV, for $2 < p_\mathrm{T} < 3$~GeV/$c$ (left) and $3 < p_\mathrm{T} < 5$~GeV/$c$ (right).}
\end{figure}

\begin{figure}[h!t]
\centering
\begin{minipage}{14pc}
\includegraphics[width=14pc]{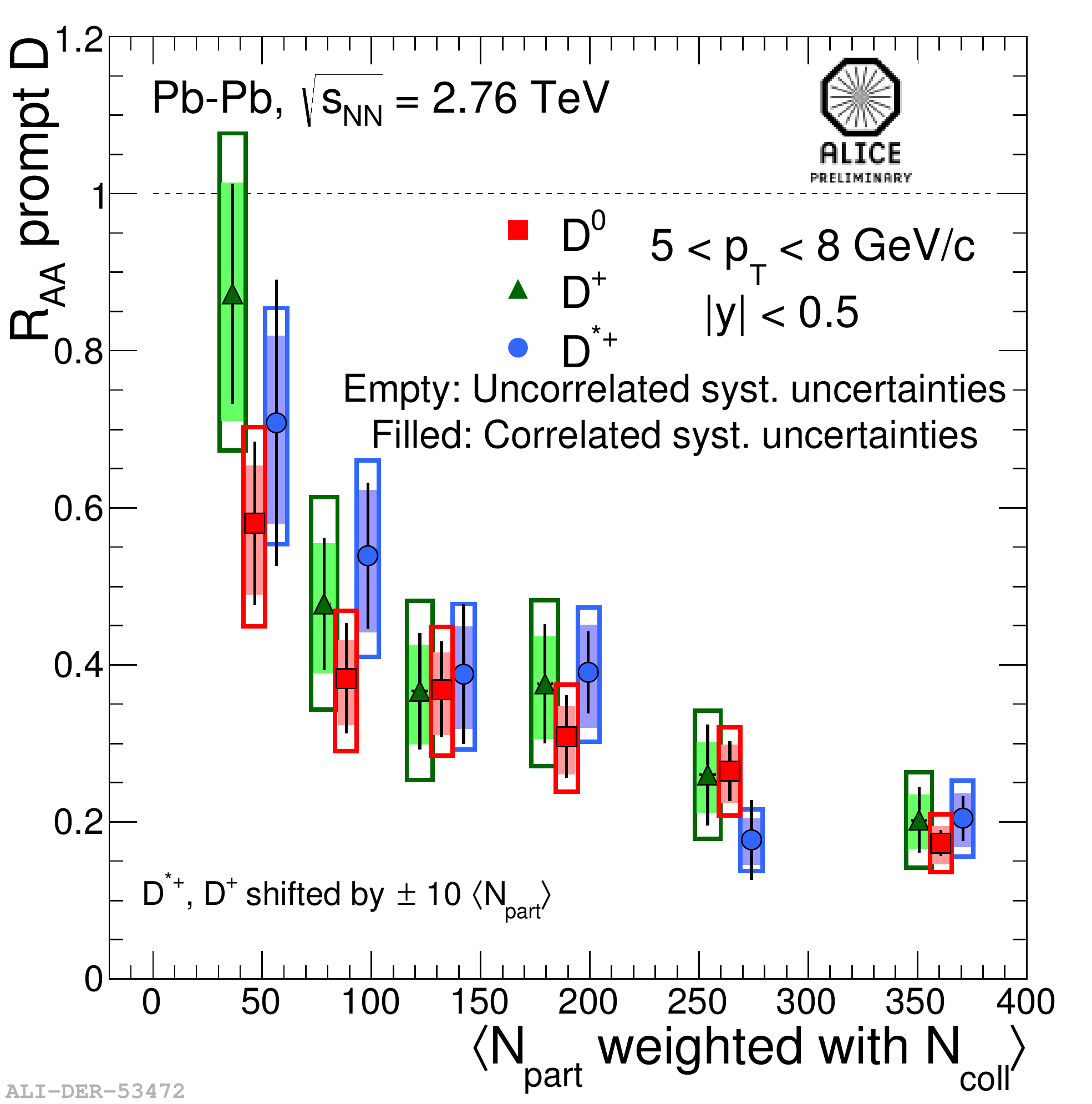}
\end{minipage}\hspace{2pc}%
\begin{minipage}{14pc}
\includegraphics[width=14pc]{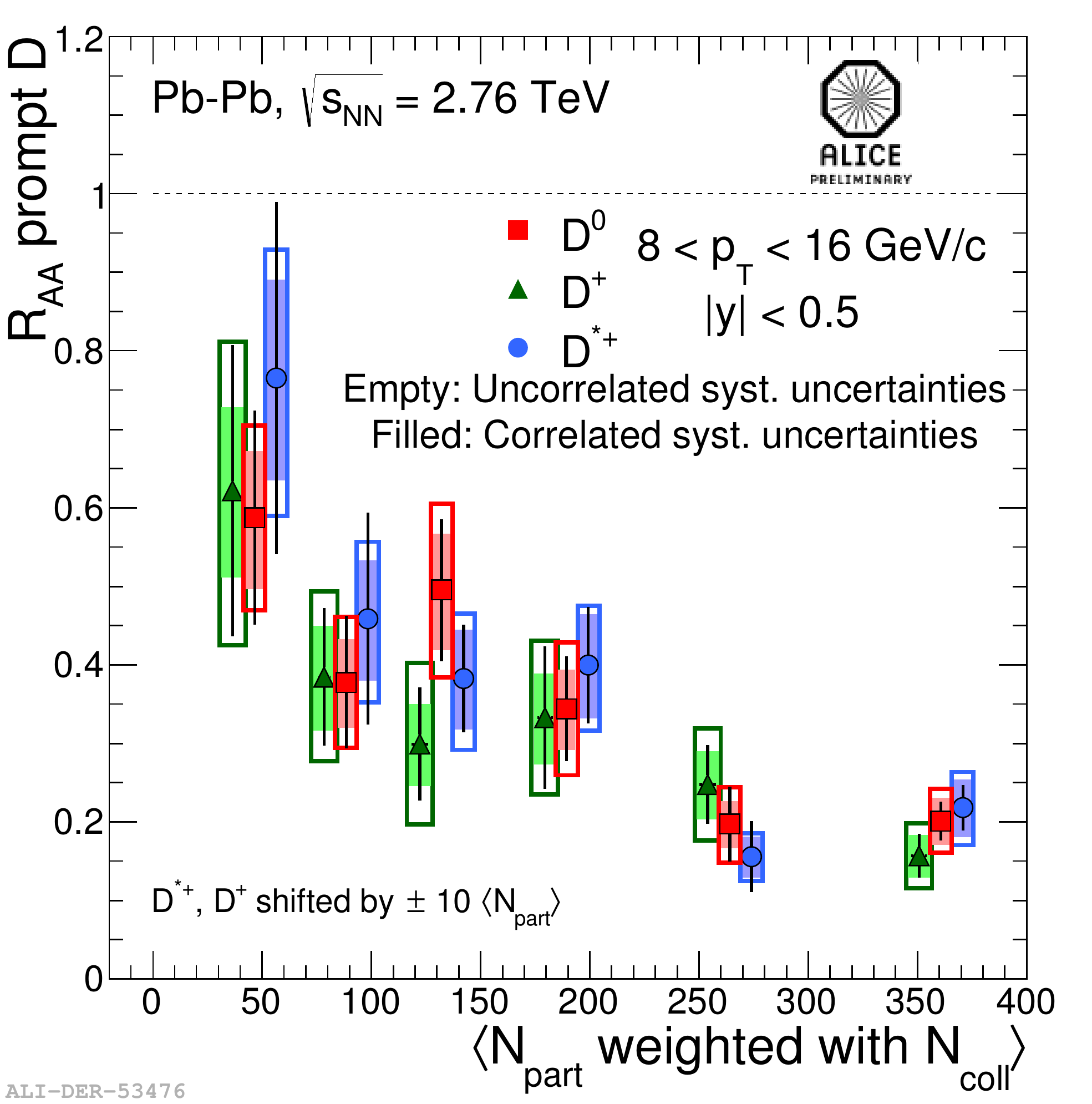}
\end{minipage} 
\caption{\label{fig:Dallraavscent1} \label{fig:Dallraavscent2}  $R_\mathrm{AA}$ as a function of centrality for D$^0$, D$^+$ and D$^{*+}$ mesons in Pb--Pb collisions at $\sqrt{s_\mathrm{NN}}=2.76$~TeV, for $5 < p_\mathrm{T} < 8$~GeV/$c$ (left) and $8 < p_\mathrm{T} < 16$~GeV/$c$ (right).}
\end{figure}

\begin{figure}[h!t]
\centering
\begin{minipage}{13.2pc}
\includegraphics[width=13.2pc]{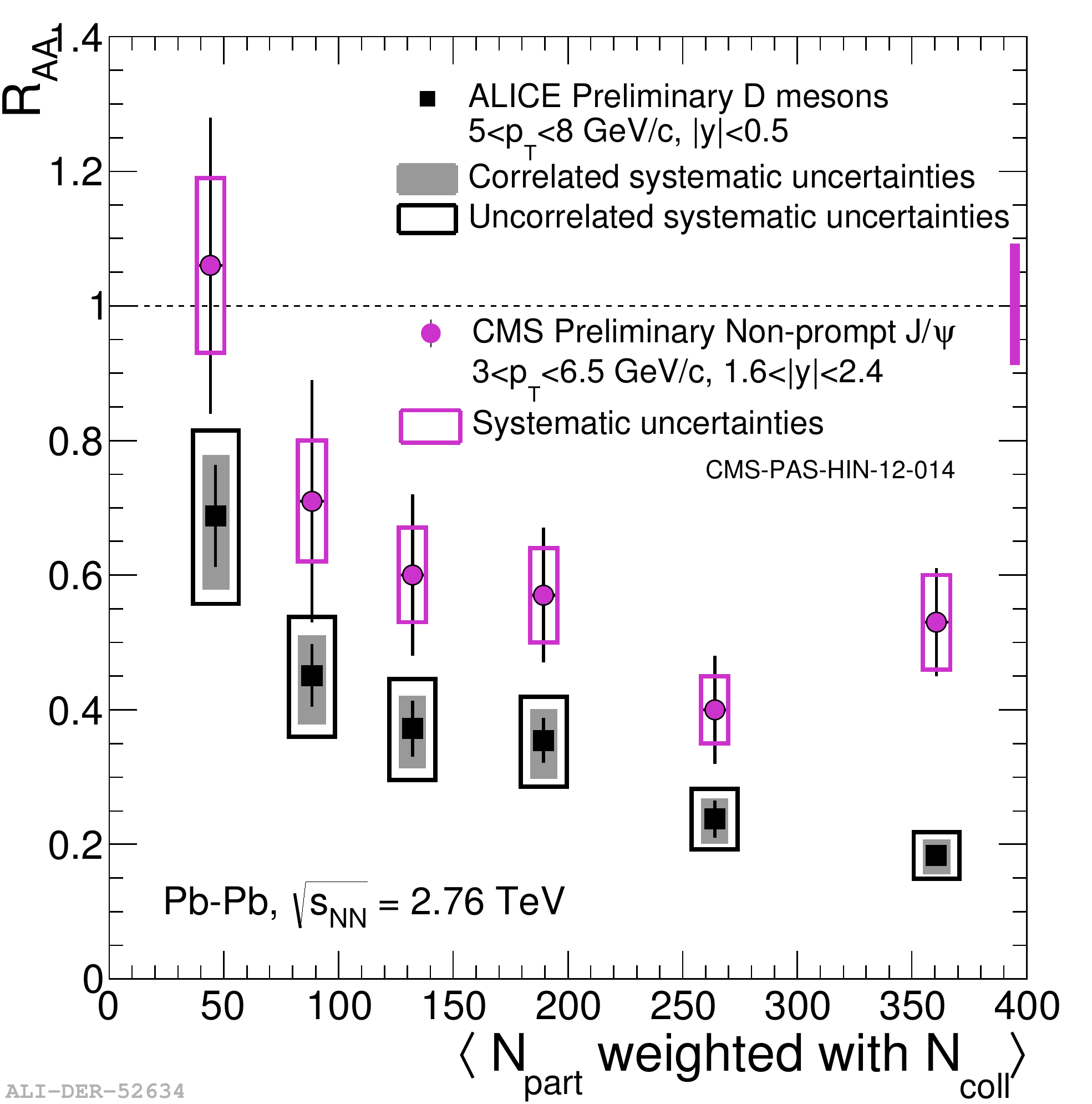}
\end{minipage}\hspace{2pc}%
\begin{minipage}{13.2pc}
\includegraphics[width=13.2pc]{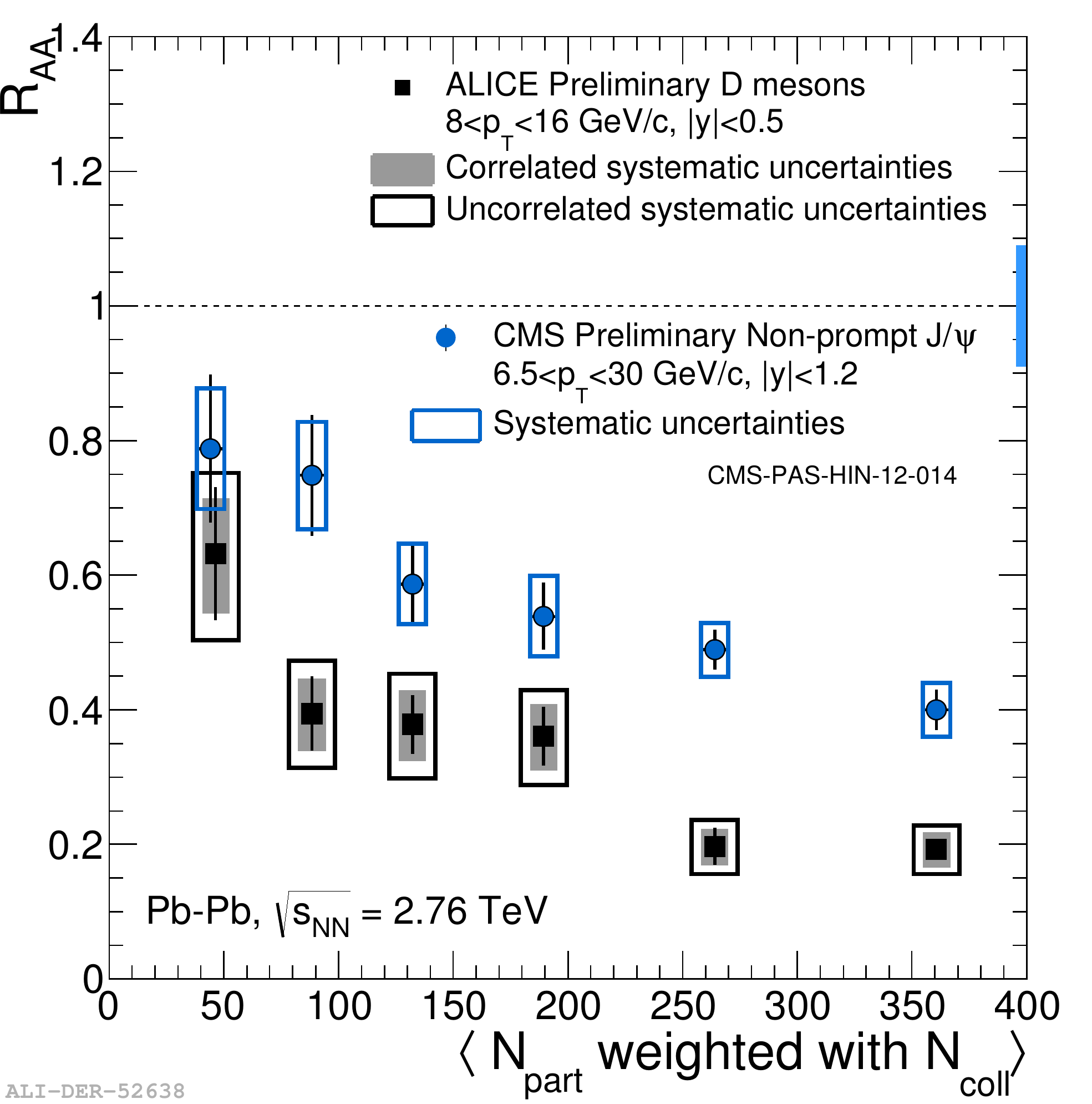}
\end{minipage} 
\caption{\label{fig:nonpromptjpsi}   Average $R_\mathrm{AA}$ as a function of centrality for D$^0$, D$^+$ and D$^{*+}$ mesons in Pb--Pb collisions at $\sqrt{s_\mathrm{NN}}=2.76$~TeV, compared with the $R_\mathrm{AA}$ of non-prompt J/$\psi$ mesons measured by the CMS collaboration in similar kinematic regions.}
\end{figure}

The centrality-dependent $R_\mathrm{AA}$ was also determined for D$^0$, D$^+$, and D$^{*+}$. The preliminary results for D$^0$ at low $p_\mathrm{T}$ are shown in figure~\ref{fig:D0raavscent1}, and at higher $p_\mathrm{T}$ for all three species in figure~\ref{fig:Dallraavscent1}. From these it can be seen that the centrality-dependent behaviour of the $R_\mathrm{AA}$ varies largely with $p_\mathrm{T}$ at low momenta, eventually stabilising at higher $p_\mathrm{T}$. In addition, there is no significant difference in the centrality dependence of the suppression between the three D-meson species.
This result was also compared with that for non-prompt J/$\psi$ mesons (i.e. J/$\psi$ particles from the decay of B mesons) measured by the CMS collaboration~\cite{cmsjpsi}. It should be noted that the CMS measurements were taken in different rapidity regions to the ALICE results. For direct comparisons to be made, the D-meson $p_\mathrm{T}$ intervals were chosen such that they represented similar kinematic regions, using kinematic simulations of B $\rightarrow$ J/$\psi$ + X. These comparisons are shown in figure~\ref{fig:nonpromptjpsi}. The results indicate that there is a greater level of suppression for charm quarks than for beauty, as expected from the predicted mass dependence discussed in section~\ref{sec:intro}.

The $v_2$ of D mesons at mid-rapidity was measured in the centrality range 30-50\%~\cite{alicepbpbv2} using the event-plane method, defined as follows:

\begin{equation}\label{eq:v2}
v_2 = \frac{1}{R_2}\frac{\pi}{4}\frac{N_\text{in-plane}-N_\text{out-of-plane}}{N_\text{in-plane}+N_\text{out-of-plane}},
\end{equation}
where the factor $\frac{1}{R_2}$ is used to correct for the finite resolution involved in determining the reaction plane. $N_\text{in-plane}$ and $N_\text{out-of-plane}$ refer to the raw yields in two azimuthal regions relative to the reaction plane: in-plane ($-\frac{\pi}{4} < \Delta\varphi < \frac{\pi}{4}$ and $\frac{3\pi}{4}<\Delta\varphi <\frac{5\pi}{4}$) and out-of-plane ($\frac{\pi}{4} < \Delta\varphi < \frac{3\pi}{4}$ and $\frac{5\pi}{4}<\Delta\varphi < \frac{7\pi}{4}$).

\begin{figure}[h!t]
\centering
\begin{minipage}{15pc}
\includegraphics[width=15pc]{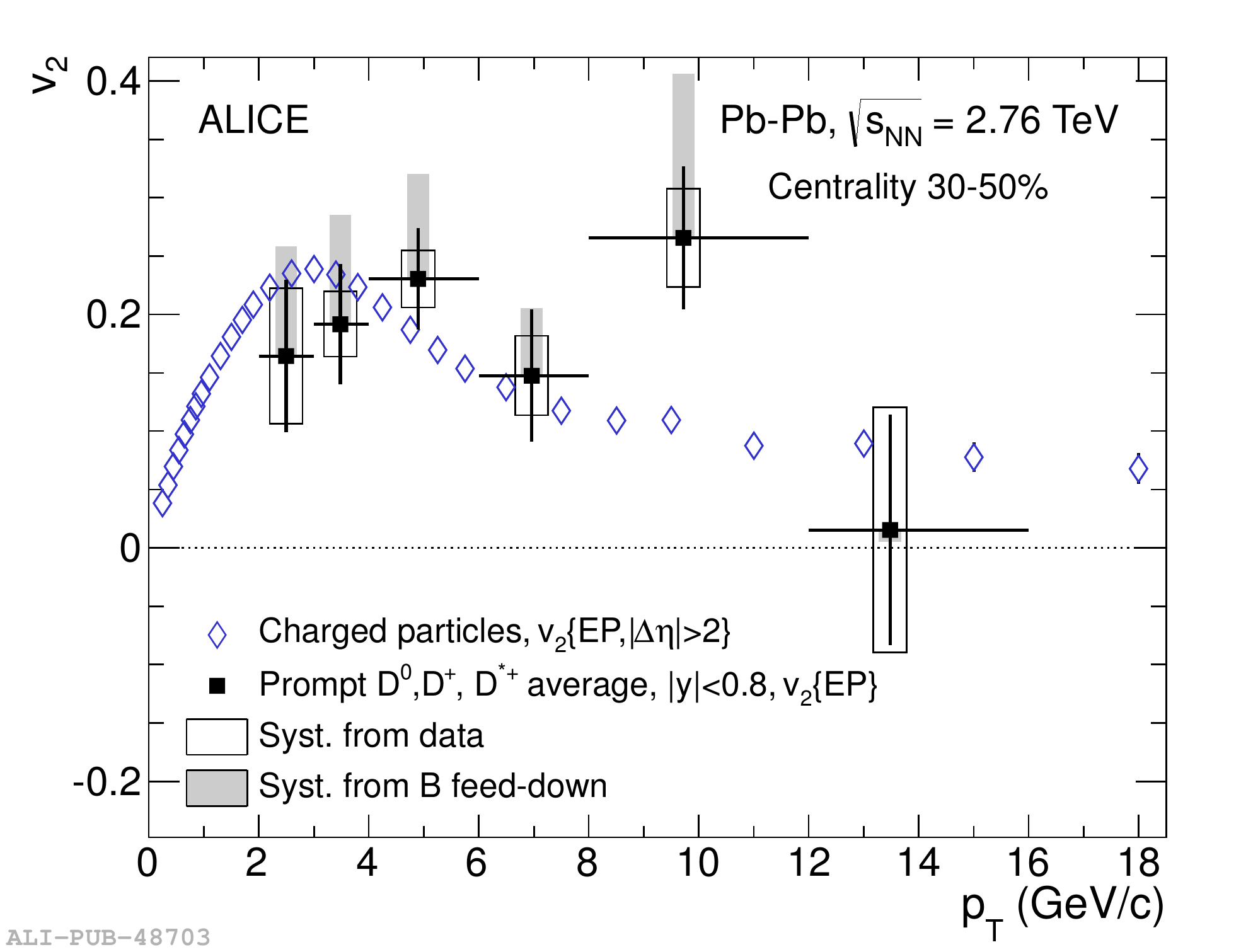}
\end{minipage}\hspace{1.5pc}%
\begin{minipage}{15pc}
\includegraphics[width=15pc]{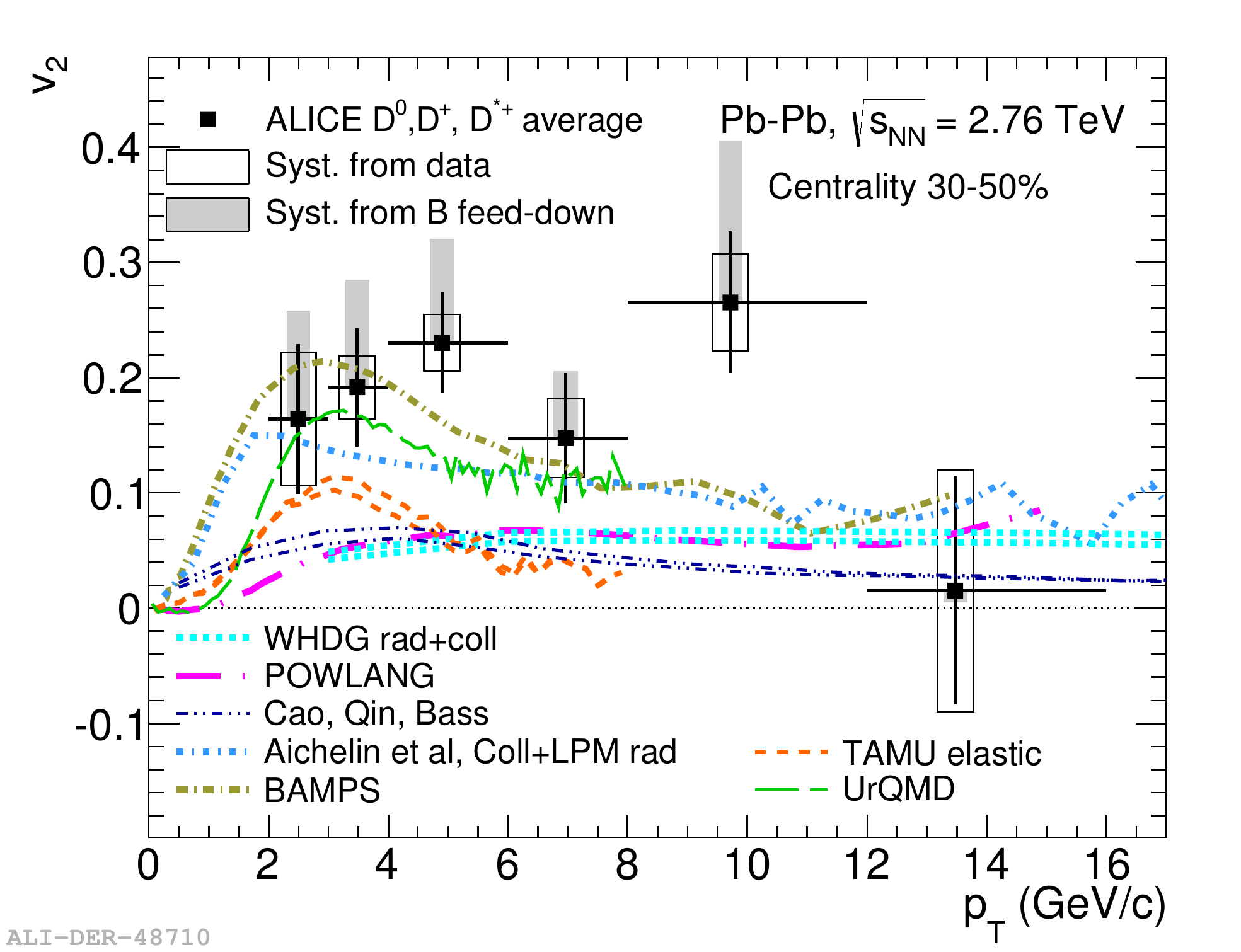}
\end{minipage} 
\caption{\label{fig:v2daverage} \label{fig:v2vsmodels} Average $v_2$ of D$^0$, D$^+$ and D$^{*+}$ mesons as a function of $p_\mathrm{T}$ for Pb--Pb collisions at $\sqrt{s_\mathrm{NN}}=2.76$~TeV in the 30-50\% centrality range, compared with the $v_2$ of charged particles (left) and with model predictions (right)~\cite{WHDG,POWLANG,caoqinbass,aichelin1,aichelin2,BAMPS,TAMU,UrQMD1,UrQMD2}.}
\end{figure}

The average $v_2$ for D$^0$, D$^+$ and D$^{*+}$ mesons is shown in figure~\ref{fig:v2daverage}, compared with the measured $v_2$ of charged particles. The D-meson $v_2$ is significantly greater than zero at low $p_\mathrm{T}$, with a greater than 5$\sigma$ significance at $2 < p_\mathrm{T} < 6$~GeV/$c$, as well as being compatible with the overall $v_2$ of charged particles (which is dominated by light-flavour hadrons). This suggests that low-momentum charm quarks take part in the collective motion of the system.

 The combination of the $R_\mathrm{AA}$ and $v_2$ measurements provides a strict constraint for the predictions given by models. By contrasting the comparison of $R_\mathrm{AA}$ to models in figure~\ref{fig:centralraavsmodels} with that of $v_2$ with model predictions in figure~\ref{fig:v2vsmodels}, it can be seen that some of the currently-available models are able to reproduce either the $R_\mathrm{AA}$ or the $v_2$ measured in experiment, however a simultaneous description of both of these results remains a challenge.

\section{Results in p--Pb collisions}  \label{sec:ppbresults}

\begin{figure}[h!t]
\centering
\begin{minipage}{14pc}
\includegraphics[width=14pc]{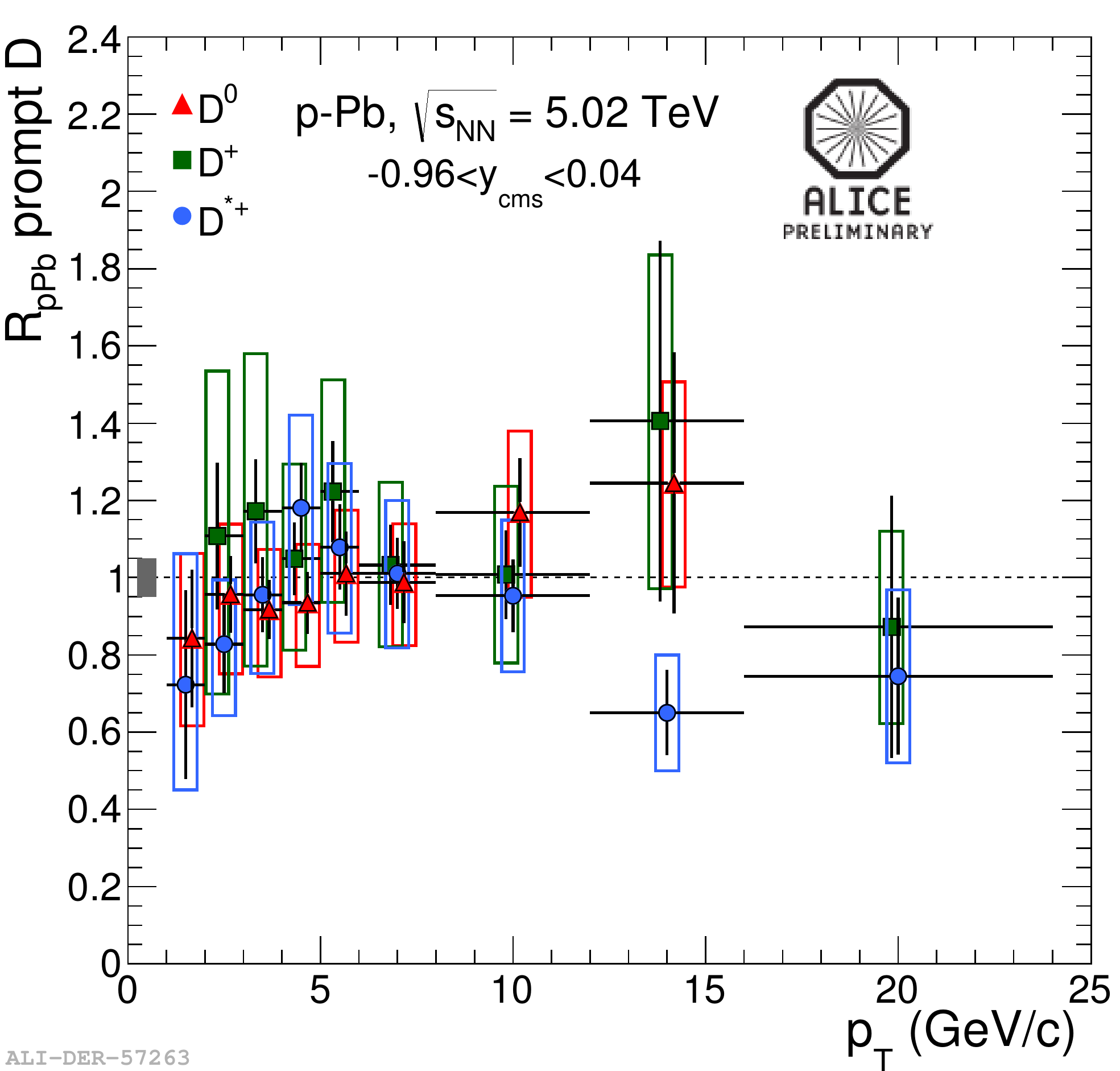}

\end{minipage}\hspace{2pc}%
\begin{minipage}{14pc}
\includegraphics[width=14pc]{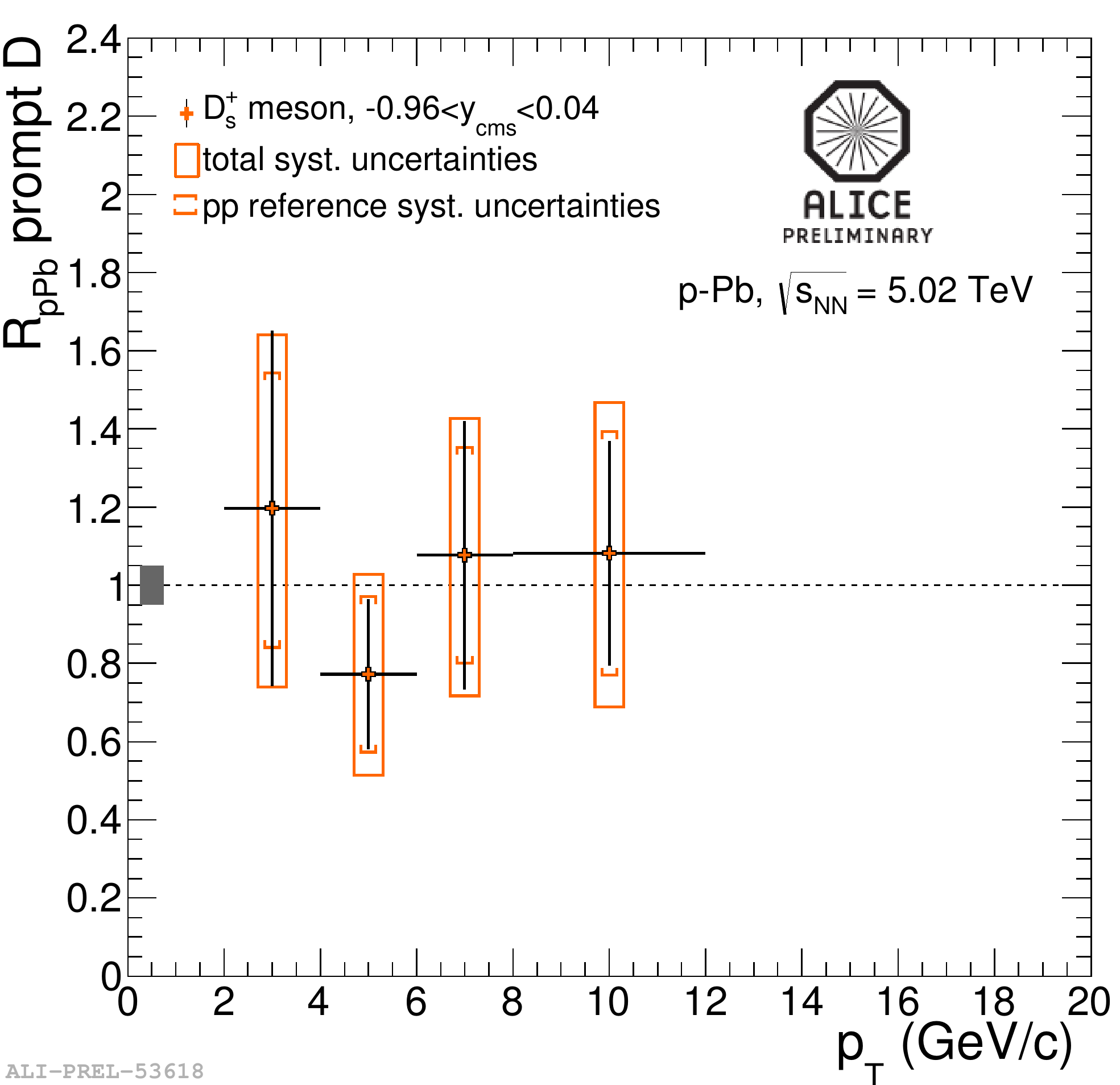}
\end{minipage} 
\caption{\label{fig:rppbd0dplusdstar}\label{fig:rppbdsplus} $R_\mathrm{pPb}$ of D$^0$, D$^+$, D$^{*+}$ (left) and D$_\mathrm{s}^{+}$ (right) mesons as a function of $p_\mathrm{T}$ in p--Pb collisions at $\sqrt{s_\mathrm{NN}}=5.02$~TeV.}
\end{figure}

The data for p--Pb collisions at $\sqrt{s_\mathrm{NN}}=5.02$~TeV used in the D-meson analysis were taken during a dedicated run at the start of 2013, just prior to start of the first Long Shutdown (LS1) of the LHC. The data were taken with a minimum-bias trigger, with an integrated luminosity of $49~\mu\mathrm{b}^{-1}$.
Once the differential cross sections were determined, they were used to compute the modification factor $R_\mathrm{pPb}$ using a similar method to the $R_\mathrm{AA}$ calculation:
\begin{equation} \label{eq:rppb}
R_\mathrm{pPb}(p_\mathrm{T}) = \frac{\mathrm{d}\sigma_\mathrm{pPb}/\mathrm{d}p_\mathrm{T}}{A\times(\mathrm{d}\sigma_\mathrm{pp}/\mathrm{d}p_\mathrm{T})},
\end{equation}
where $\sigma_\mathrm{pPb}$ is the cross section in p--Pb collisions, $\sigma_\mathrm{pp}$ is the cross section in pp collisions at the same centre-of-mass energy, and $A$ is the atomic mass number of the Pb nucleus. The preliminary $R_\mathrm{pPb}$ measurements for D$^0$, D$^+$, D$^{*+}$, and  D$_\mathrm{s}^+$ mesons are shown in figure~\ref{fig:rppbd0dplusdstar}. The measurements for all D-meson species, including D$_\mathrm{s}^+$, are compatible with one another, as well as being consistent with unity within uncertainties. 

\begin{figure}[h!t]
\centering
\begin{minipage}{14pc}
\includegraphics[width=14pc]{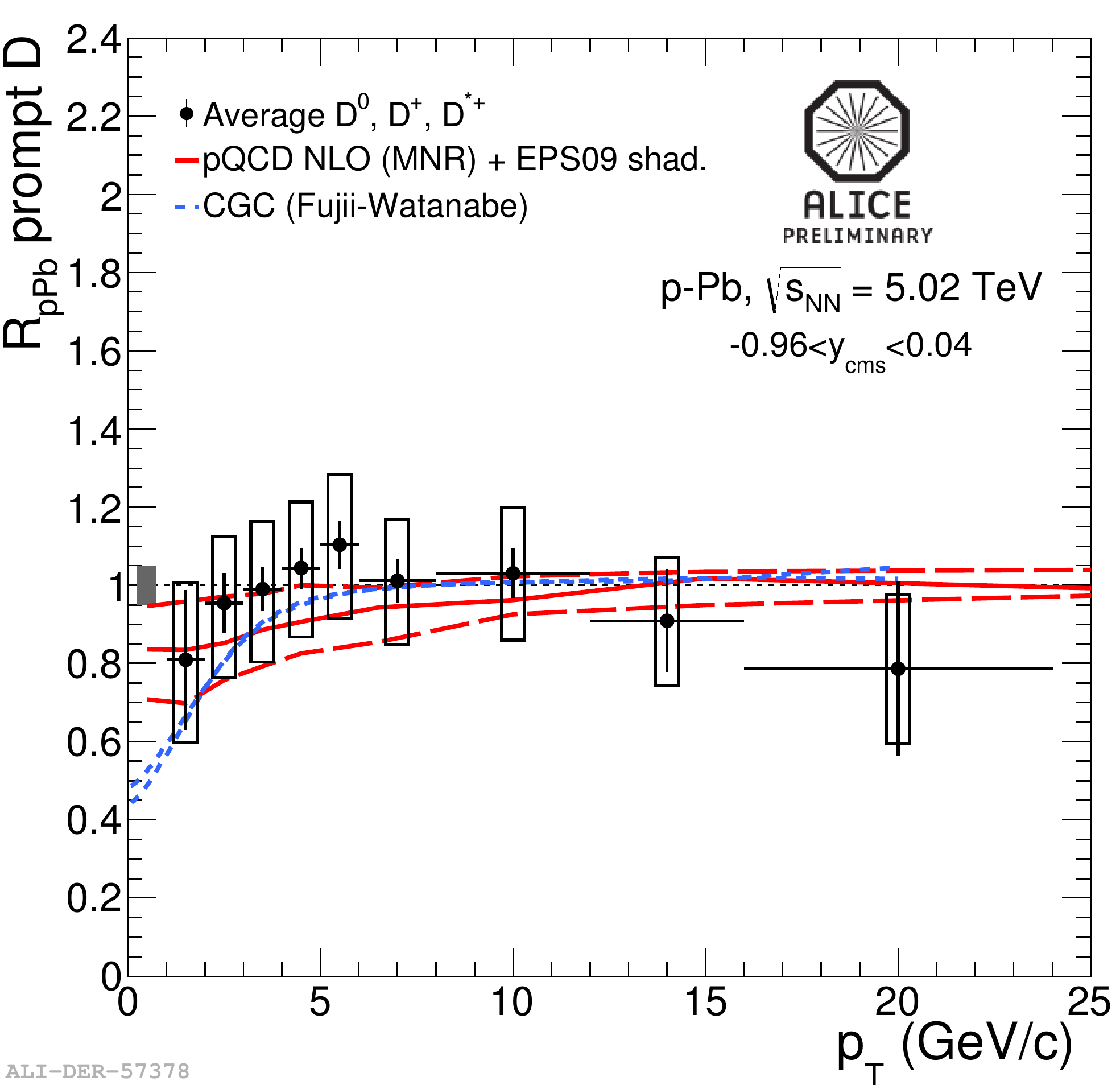}

\end{minipage}\hspace{2pc}%
\begin{minipage}{14pc}
\includegraphics[width=14pc]{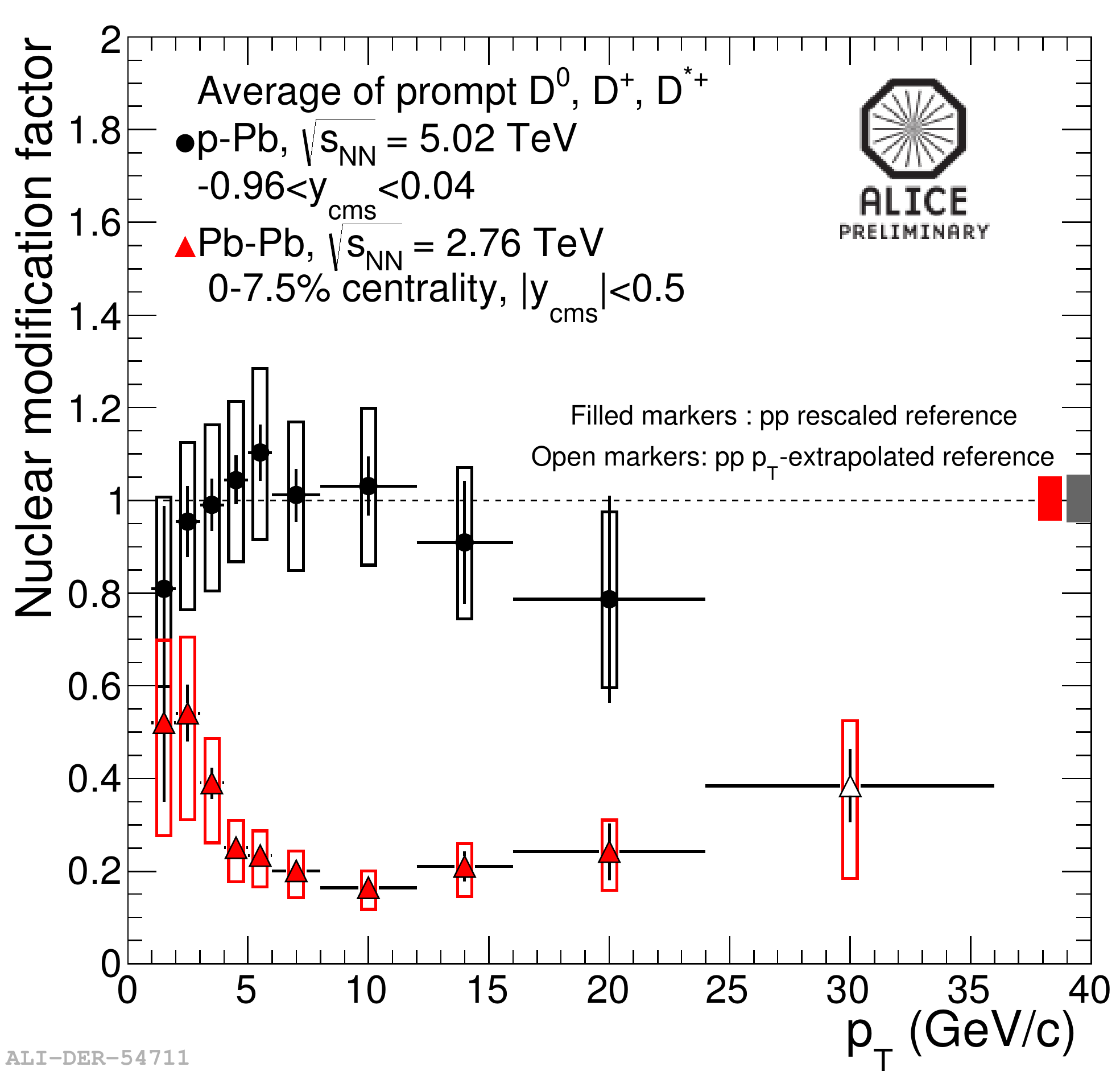}
\end{minipage} 
\caption{\label{fig:rppbmodels}\label{fig:rppbraa} Average $R_\mathrm{pPb}$ of D$^0$, D$^+$ and D$^{*+}$ mesons as a function of $p_\mathrm{T}$ at $\sqrt{s_\mathrm{NN}}=5.02$~TeV, compared with models~\cite{MNR,EPS09} (left) and $R_\mathrm{AA}$ (right)}
\end{figure}

The $R_\mathrm{pPb}$ measurements for D$^0$, D$^+$ and D$^{*+}$ mesons were averaged and compared with results from two models, namely next-to-leading order pQCD calculations using the MNR framework~\cite{MNR} with EPS09 for the description of the nuclear PDFs~\cite{EPS09}; and the Colour Glass Condensate (CGC) framework~\cite{CGC}. This comparison is shown in figure~\ref{fig:rppbmodels}. It can be seen that, within uncertainties, the experimental results  do not favour one prediction over the other.

Finally, in figure~\ref{fig:rppbraa}, we also compare the average $R_\mathrm{pPb}$ with the average $R_\mathrm{AA}$ measured for D mesons, which was discussed in section~\ref{sec:pbpbresults}. We find that, although the centre-of-mass energies of the two systems differ, there is a clear suppression in Pb--Pb collisions compared with p--Pb for $p_\mathrm{T} > 4$~GeV/$c$. This indicates that final-state effects such as partonic energy loss in the medium have a far more dominant role in the suppression of D-meson production at higher $p_\mathrm{T}$ than initial-state effects such as nuclear shadowing.

\section{Conclusions and Outlook}
The reconstructed decays of D mesons were measured at mid-rapidity in pp, Pb--Pb and p--Pb collisions. The pp measurements, which were used as reference spectra for the heavy-ion results, were found to be in agreement with pQCD calculations, and extrapolations of the cross sections to full phase space were consistent with the trends exhibited by both experiment and theory across three orders of magnitude in energy.

 In Pb--Pb collisions, it was determined that the level of suppression experienced by non-strange D mesons is similar between species, with a significant suppression exhibited at high $p_\mathrm{T}$. This suppression can be understood as an indication of in-medium parton energy loss. Further study will be required before conclusions can be drawn as to whether there is an enhancement of the D$_\mathrm{s}^+$ yield relative to non-strange D mesons at low $p_\mathrm{T}$. In addition, as expected due to the dead-cone effect, the $R_\mathrm{AA}$ of D mesons was found to be lower than that of B mesons in the range $5<p_\mathrm{T}<16$~GeV/$c$. Finally, a significantly non-zero $v_2$ for D mesons was measured at low $p_\mathrm{T}$, suggesting that charm quarks take part in the collective motion of the medium. The results from p--Pb collisions show the $R_\mathrm{pPb}$ to be consistent with unity, as well as being compatible with models of initial-state effects. This suggests that the suppression visible at high $p_\mathrm{T}$ in Pb--Pb collisions primarily occurs due to hot nuclear matter effects rather than initial-state effects.

The rapidity and multiplicity dependence of D-meson production in p--Pb collisions are currently being studied. In 2015, once the current Long Shutdown of the LHC is complete, ALICE will resume taking data at a higher $\sqrt{s_\mathrm{NN}}$, allowing further refinement of the results with an expected factor of 10 increase of statistics prior to the start of the second Long Shutdown.

\section*{References}
\bibliographystyle{iopart-num}
\bibliography{hptproc}

\end{document}